\documentclass[twocolumn]{aastex631}

\newcommand{\ars}{\langle R_{\rm s} \rangle}

\newcommand{\nH}{n_{\rm H,\infty}}
\newcommand{\cmq}{{\rm cm}^{-3}}

\newcommand{\hii}{\rm H\,{\textsc{ii}}}

\newcommand{\msun}{{\rm M_{\sun}}}

\newcommand{\mbh}{M_{\rm BH}}

\newcommand{\enstatite}{{\rm MgSiO_3}}

\shorttitle{Dusty gas accretion of seed black holes}
\shortauthors{Park et al.}

\begin{document}

\title{Accelerated growth of seed black holes by dust in the early universe}

\correspondingauthor{KwangHo Park}
\email{kwangho.park@physics.gatech.edu}

\author[0000-0001-7973-5744]{KwangHo Park}
\affiliation{School of Physics, Georgia Institute of Technology, Atlanta,
GA 30332, USA}

\author[0000-0001-6246-2866]{Gen Chiaki}
\affiliation{School of Physics, Georgia Institute of Technology, Atlanta,
GA 30332, USA}
\affiliation{Astronomical Institute, Graduate School of Science, Tohoku University, Aoba, Sendai 980-8578, Japan}

\author[0000-0003-1173-8847]{John H. Wise}
\affiliation{School of Physics, Georgia Institute of Technology, Atlanta,
GA 30332, USA}

\begin{abstract}
We explore the effect of dust on the growth of seed black holes (BHs) in the early universe. Previous 1D radiation-hydrodynamic (RHD) simulations show that increased radiation pressure on dust further suppresses the accretion rate than the case for the chemically pristine gas. Using the Enzo+Moray code, we perform a suite of 3D RHD simulations of accreting BHs in a dusty interstellar medium (ISM). We use the modified Grackle cooling library to consider dust physics in its non-equilibrium chemistry. The BH goes through an early evolutionary phase, where ionizing BH radiation creates an oscillating \ion{H}{2} region as it cycles between accretion and feedback. As the simulations proceed, dense cold gas accumulates outside the ionized region where inflow from the neutral medium meets the outflow driven by radiation pressure. In the late phase, high-density gas streams develop and break the quasi-spherical symmetry of the ionized region, rapidly boosting the accretion rate. The late phase is characterized by the coexistence of strong ionized outflows and fueling high-density gas inflows. The mean accretion rate increases with metallicity reaching a peak at Z$\sim$0.01--0.1\,$Z_\odot$, one order of magnitude higher than the one for pristine gas. However, as the metallicity approaches the solar abundance, the mean accretion rate drops as the radiation pressure becomes strong enough to drive out the high-density gas. Our results indicate that a dusty metal-poor ISM can accelerate the growth rate of BHs in the early universe, however, can stun its growth as the ISM is further enriched toward the solar abundance.

\end{abstract}






\section{Introduction}

Bright quasars powered by supermassive black holes (SMBHs) shine at the edge of the observable universe at redshifts greater than six.  Only after less than a billion years, these distant beacons already have masses above $10^9\,\msun$ \citep[e.g.,][]{Fan:2001,Willott:2003, Mortlock:2011, Wu:2015, Banados:2018}, comparable to the most massive BHs in the local universe.  Their rapid growth during the early universe raises fundamental astrophysical questions about their formation, evolution, and connection with galaxy formation \citep{Kormendy:2013}.

Numerical simulations suggest that intermediate-mass black holes (IMBHs) in the 
range of $\mbh=10^2$--$10^5~\msun$ might be able to form in the early universe. 
This is possible due to the unique physical conditions which produce much 
heavier seeds than the typical stellar remnants in the local universe. The scenarios include the remnants of the 
deaths of very massive metal-free (Pop\,III) stars \citep{BrommCL:99,AbelBN:00,
MadauR:01}, a dense stellar cluster collapse \citep{Devecchi:2009, Davies:2011,
Lupi:2014, Katz:2015,Boekholt:2018, Reinoso:2018}, and a direct collapse of 
chemically pristine gas \citep{BegelmanVR:06,ChoiSB:13,YueFSXC:14, Regan:2017}.


However, even the direct collapse BH (DCBH) scenario which predicts the most massive seeds $\sim 10^5\,\msun$, 
it is still challenging to make those seeds grow to $\gtrsim 10^9\,\msun$ 
quasars by $z \sim 6$. Radiation hydrodynamics (RHD) simulations show that BHs can 
grow rapidly only under special conditions \citep[e.g.,][]{Pacucci:2017}. This is because with rapid growth comes strong radiative feedback, where UV and X-ray photons heat and ionize the surrounding interstellar medium (ISM), creating an $\hii$ region around an accreting BH \citep{MiloCB:09, ParkR:11, ParkR:12}. 
This hot bubble filled with low-density gas regulates the gas supply to the 
BHs. Due to the cycles between accretion and feedback, the accretion behavior 
is highly oscillatory creating constantly fluctuating $\hii$ regions. That 
introduces turbulence in the ISM, however \citet*[][hereafter, PWB17]
{ParkWB:2017} find that the the thermal energy dominates the turbulent kinetic 
energy. 

This {\it feedback-dominated} accretion occurs when average size of the $\hii$ region $\ars$ exceeds the Bondi radius $r_{\rm B}=GM_{\rm BH}/c_{\rm s}^2$. Here, $M_{\rm BH}$ is the BH 
mass, and $c_{\rm s}$ is the sound speed of the gas. By definition, the Bondi radius is the scale within which the gravitational energy of the BH dominates over 
the thermal energy of the gas, and $\ars$ is another scale that the BH 
radiation heats and ionizes. Note that this scale is by far larger than the 
accretion disk scale and regulates gas supply to the BH from galactic scales.
The condition for feedback-limited regime 
assuming the gas temperature $T_\infty =10^4$\,K and the standard 
radiative efficiency $\eta =0.1$, 
can be simplified as the product of the 
BH mass and the gas density as $\mbh\,\nH \le 10^9\,\msun \cmq$ where $\nH$ is the hydrogen number density of the gas. 
The mean accretion rate can be expressed assuming the 
spectral index is $\alpha_{\rm spec}=1.5$  as the following \citep{ParkR:12}
\begin{equation}
   \langle \dot{M}_{\rm BH} \rangle \sim 6\times 10^{-6} 
   M_{\rm BH, 4}^2\,
   n_{\rm H, \infty, 3}\,
   T_{\infty, 4}\,
   \,{\rm \msun}{\rm yr}^{-1}, 
   \label{eq:pressureequilibrium}
\end{equation}
where $M_{\rm BH, 4} = {\mbh}/(10^4\,\msun)$, $n_{\rm H, \infty, 3}=\nH/(10^3\,\cmq)$, and $T_{\infty, 4}={T_\infty}/(10^4\,{\rm K})$. Note that the mean accretion rate is proportional to the gas pressure for a given BH mass (i.e., $\langle \dot{M}_{\rm BH} \rangle  \propto n_{\rm H, \infty}\, T_{\infty}$).

In more extreme conditions, the accretion flow makes a transition to {\it 
feeding-dominated} regime \citep{PacucciVF:2015}. When $\ars \le r_B$ which is translated as $\mbh 
\nH \ge 10^9\,\msun\cmq$ assuming the same $T_\infty$ and $\eta$, the ionization front becomes unstable and is accepted as the condition for {\it hyperaccretion} \citep
{ParkRDR:14a,Inayoshi:2016}. There have been several 
works to study the geometry of the hyperaccretion flow such as anisotropic radiation 
\citep{Sugimura:2016} or biconical-dominated accretion flow (BDAF) and 
decretion disk \citep{Park:2020}.

Several scenarios have been suggested for exponential accretion \citep[e.g., ][]
{AlexanderN:2014}. Among them, it might be worth noting that some works pointed 
out the possible connection with the early assembly of stellar bulge \citep
{Park:2016,Inayoshi:2021}. \citet{Park:2016} pioneered the possibility of 
bulge-driven growth of seed BHs. This work predicts that there exists a critical 
bulge mass of $\sim 10^6\,\msun$ above which the rapid growth of seed BHs is 
triggered no matter of the central BH masses. This mechanism could favor DCBHs as the seeds of high-redshift quasars because immediately their formation, efficient star formation is triggered by positive radiative feedback \citep{Barrow:2018}. This newly formed nuclear star cluster would then aid in subsequent growth as the young stellar bulge component increases the 
effective accretion radius supplying more gas from large scale to the seed BHs. 
Recently, \citet{Inayoshi:2021} perform asymmetric 2D simulations, suggesting 
obese BHs \citep[OBGs;][]{Agarwal:2013} as the progenitor of the high-redshift 
quasars, which are significantly overmassive compared to the observed bulge-to-BH mass relation \citep{Kormendy:2013}.


Another critical uncertainty in the growth of seed BHs, which is the main focus of the current study, is the role of dust. Observation 
discovers that the early universe is rapidly contaminated with dust by supernovae. For example, a star-forming galaxy A1689-zD1 at $z > 7$ that is 
heavily enriched in dust, and the dust-to-gas ratio is close to that of the 
Milky Way \citep{Watson:2015}. Atacama Large Millimeter Array (ALMA) 
observations also detected dust-contaminated quasars at $6.0 \lesssim z < 6.7$ 
with the estimated dust masses of $M_{\rm d} = 10^7$--$10^9$\,$\msun$ \citep
{Venemans:2018}. Cosmological simulations also predict that the early universe 
is rapidly contaminated by the ongoing star-formation. A single pair-instability supernova can enrich a halo of $10^7$\,$\msun$ to a metallicity of $10^{-3}Z_\sun$, triggering a transition from Pop~III to Pop~II star 
formation \citep{Bromm:2003, Wise:2012}. \citet{Yajima:2015} show that the amount of dust in an overdense 
region of massive galaxies can reach the solar neighborhood level even at 
$z\gtrsim 6$. Interestingly, the contamination of the dust is closely associated with the 
bulge-driven growth of the BHs discussed above since the assembly of the 
stellar bulge leads to metal enrichment of the ISM around the BHs. Recently, 
\citet{Ishibashi:2021} explores the parameter space for dust photon trapping 
(DPT) which might lead to the supercritical BH growth.

Previous work by \citet{Yajima:2017} was performed using 1D RHD simulations for a relatively short period of time compared to the time scale in the current study. They discover that the radiation pressure on dust is the important parameter compared to the dust attenuation effect. The average accretion rate is lower than the case of the primordial gas. The spectral energy distribution (SED) is calculated from the dust thermal emission and they suggest that the flux ratio between $\lambda \lesssim 20 \mu$m and $\gtrsim 100\,\mu$m shows a close relation to the Eddington ratio.

\begin{table*}
   \begin{center}
   \caption{Simulation Parameters} 
   \begin{tabular}{ccccc}
   \hline
   \hline
   ID   &   $Z/Z_\odot$  & $\tau_{\rm tran}$ (Myr) & $\langle \dot{M}_{\rm early} \rangle$ ($\msun\,{\rm yr}^{-1}$) 
   &  $\langle \dot{M}_{\rm late} \rangle $ ($\msun\,{\rm yr}^{-1}$)  \\
   \hline
   {Zm0} & $10^{0}$      & 7.2 & $(1.7_{-1.6}^{+6.1}) \times 10^{-7}$ &  $(3.7_{-3.4}^{+10.0}) \times 10^{-6} $ \\
   {Zm05} & $10^{-0.5}$  & 8.2 & $(1.2_{-1.0}^{+3.0}) \times 10^{-7}$ &  $(1.1_{-0.99}^{+1.2}) \times 10^{-5} $ \\
   {Zm1} & $10^{-1}$     & 7.5 & $(1.9_{-1.8}^{+5.8}) \times 10^{-7}$ &  $(5.3_{-5.2}^{+11.0}) \times 10^{-5} $ \\
   {Zm2} & $10^{-2}$     & 5.2 & $(3.8_{-3.6}^{+10.0}) \times 10^{-7}$ &  $(7.5_{-4.5}^{+5.8}) \times 10^{-5} $ \\
   {Zm3} & $10^{-3}$     & 5.3 & $(3.9_{-3.7}^{+13.0}) \times 10^{-7}$ &  $(3.5_{-2.2}^{+11.0}) \times 10^{-5} $ \\
   {Zm4} & $10^{-4}$     & 4.7 & $(4.2_{-3.9}^{+13.0}) \times 10^{-7}$ &  $(1.8_{-0.95}^{+1.3}) \times 10^{-5} $ \\
   {Zm6} & $10^{-6}$     & 4.7 & $(3.8_{-3.5}^{+11.0}) \times 10^{-7}$ &  $(2.0_{-0.75}^{+0.71}) \times 10^{-5} $ \\
 

   \hline
   \end{tabular}
   \tablecomments{$\mbh=10^4\,\msun$ and $\nH=10^3\,{\rm cm}^{-3}$, $D_{\rm box}$=80\,pc, and $\Delta D_{\rm min}$=0.625\,pc for all runs. We assume $Z_\odot=0.02$. The upper and lower errors for each accretion rate represent the 5\%  and 95\% ($2\sigma$) percentiles.}
   \label{table:para}
   \end{center}
\end{table*}


Future multi-messenger observations of the cosmic dawn will reveal how the first luminous structures co-evolve with massive BHs.  Some examples include radiation from the first galaxies and BHs with JWST (James Webb Space Telescope) \citep[e.g.,][]{Natarajan:2017}, X-ray observations of actively accreting BHs by ATHENA (Advanced Telescope for 
High-ENergy Astrophysics), and gravitational waves from merging BHs by LISA
(Laser Interferometer Space Antenna).

This paper builds upon the previous 1D work by \citet{Yajima:2017} to investigate the effect of the dust contamination onto BHs growth in the early universe using 3D RHD simulations with various metallicities. We emphasize that we investigate the case of the feedback-dominated regime with dust physics included, where the growth of the 
BH is highly suppressed for chemically pristine gas. We find that the dust 
cooling lowers the mean accretion rate due to lower thermal pressure in the early phase, however, boosts the accretion rates when the spherical symmetry is broken and the dense stream of gas develops an inflow channel. We also find that 
metallicity affects the mean accretion rates. In Section~\ref{sec:method}, we 
describe the numerical simulations. In Section~\ref{sec:results} we present the simulation results. We discuss the results in Section~\ref{sec:discussion} and summarize in Section~\ref{sec:summary}.

\section{Method}
\label{sec:method}

\subsection{Radiation hydrodynamic simulations}

We perform 3D RHD simulations to study the role of dusty gas on the growth seed BHs under the influence of radiative feedback. We use 
the adaptive mesh refinement (AMR) code {\it Enzo} equipped with the {\it Moray} package to solve radiative transfer equations \citep{Wise:2011,Bryan:2014,Brummel-Smith:2019}. 


We use the chemistry and cooling library {\it Grackle} \citep{Smith:2017} that has been modified by adding metal and dust grain chemistry into the primordial chemical network of {\it Grackle}, while considering the growth of dust grains \citep{Chiaki:2019, Chiaki:2021}.
The modified version of {\it Grackle} calculates all relevant chemical reactions and radiative cooling and heating 
self-consistently.\footnote{The version of {\sc grackle} used in this work is available at  \url{https://github.com/genchiaki/grackle/tree/metal-dust}.}
It includes a comprehensive chemical network of 92 reactions of 35 gas-phase species: ${\rm H}$, ${\rm H^+}$, ${\rm H_2}$, ${\rm H^-}$, ${\rm H_2^+}$, ${\rm e^-}$, ${\rm D}$, ${\rm D^+}$, ${\rm D^-}$, ${\rm HD}$, ${\rm HD^+}$, ${\rm He}$, ${\rm He^+}$, ${\rm He^{2+}}$, ${\rm HeH^+}$, ${\rm C^+}$, ${\rm C}$, ${\rm CH}$, ${\rm CH_2}$, ${\rm CO^+}$, ${\rm CO}$, ${\rm CO_2}$, ${\rm O^+}$, ${\rm O}$, ${\rm OH^+}$, ${\rm OH}$, ${\rm H_2 O^+}$, ${\rm H_2O}$, ${\rm H_3O^+}$, ${\rm O_2^+}$, ${\rm O_2}$, ${\rm Mg}$, ${\rm Si}$, ${\rm SiO}$ and ${\rm SiO_2}$.
In this study, we also trace the evolution of 2 grain species abundances: 
silicate (enstatite, $\enstatite$) and 
graphite (${\rm C}$).
We calculate the cooling by dust thermal emission, H$_2$ formation on grain surfaces, and continuum opacity of the silicate and graphite.
We describe the detail of the dust physics in Section \ref{sec:dust_modeling}.

Table~\ref{table:para} shows the list of simulations. We fix the BH mass $\mbh=10^4\,
\msun$ and gas density $\nH=10^3\,\cmq$. We also use the same domain size of $(80\,{\rm pc})^3$ with the resolution of $\Delta D_{\rm min}=0.625\,$pc. The IDs in the first column of Table~\ref{table:para} show the metallicities as Zmn where $n$ is $Z/Z_\sun = 10^{-n}$. The 3rd column is the time $\tau_{\rm tran}$ from an early oscillatory phase to a late stable phase with higher accretion rates, and the 4th and 5th columns are the mean accretion rates for each phase, respectively.

\subsection{Accretion rates of dusty gas}

In the simulation, we follow the accretion flows in a dusty ISM around a seed BH.  To quantify their behavior and evolution, we compare our results to the standard Bondi solution. The thermal state of the accretion flow is affected by adiabatic gas heating, shocks, photoionization, recombination, and dust heating and sublimation. To calculate the bolometric accretion luminosity, we use the Eddington-limited Bondi recipe that sources the radiative transfer equation. The Bondi radius is normalized as
\begin{equation}
   r_{\rm B} = \frac{G\mbh}{c_\infty^2}
   \simeq 5.2\,{\rm pc} \left(\frac{\mbh}{10^4\,\msun}\right)
   \left(\frac{T_\infty}{10^3\,{\rm K}}\right)^{-1},
   \label{eq:bondiradius}
\end{equation}   
where $c_\infty$ is the sound speed of the gas outside the ionized region that is not affected by the BH radiation or gravity and $T_\infty$ is the relevant gas temperature. We normalize the gas temperature of $T_\infty =10^3\,$K due to dust cooling and adopt $c_\infty = 9.1\,{\rm km\,s^{-1} (T_\infty /10^4\,K)^{1/2}}$ assuming isothermal gas with a mean molecular weight of 1. The Bondi accretion rate is defined as
\begin{equation}
   \dot{M}_{\rm B}=4\pi\,\lambda_{\rm B}\,r_{\rm B}^2\,\rho_\infty\,c_\infty ,
   \label{eq:bondiacc}
\end{equation}   
where the dimensionless accretion rate $\lambda_{\rm B}$ depends on the equation of state and $\rho_\infty$ is the gas density far from the BH. Since the gas is under the influence of the BH radiation, we use the gas density $\rho_{\rm HII}$ and sound speed $c_{\rm HII}$ of the cell that contains the BH, arriving at
\begin{equation}
   \dot{M}_{\rm BH}=4\pi\,\lambda_{\rm B}\,r_{\rm B, HII}^2\,\rho_{\rm HII}\,c_{\rm HII},
   \label{eq:acc}
\end{equation}   
where $r_{\rm B, HII}=G\mbh/c_{\rm HII}^2$ is the local Bondi radius under the influence of radiation. By using the ratio between the temperatures of the ionized and neutral region 

\begin{equation}
   \delta T_{\rm d} = \frac{T_{\rm HII}}{T_\infty}\simeq \frac{\rho_\infty}{2\rho_{\rm HII}}, 
\end{equation}
where the pressure equilibrium across the ionization front is applied and the ionization fraction in the $\hii$ region is $\sim 1$ as $2\rho_{\rm HII} T_{\rm HII}= \rho_\infty T_\infty$. Then the BH accretion rate is 
\begin{eqnarray}
   \langle \dot{M}_{\rm BH, d} \rangle &=& 4\pi\,\lambda_{\rm B}\,\frac{G^2 \mbh^2}{c_{\rm s}^3} \rho_\infty\,
   \left( \frac{c_{\rm s}}{c_{\rm HII}} \right)^3 
   \left( \frac{\rho_{\rm HII}}{\rho_\infty} \right) \nonumber \\
   &=& \frac{1}{2} \dot{M}_{\rm B} (\delta T_{\rm d})^{-5/2}, 
   \label{eq:mdot}
\end{eqnarray}
where $\dot{M}_{\rm BH, d}$ is $1.8\times 10^{-6}$ times of the Bondi rate $\dot{M}_{\rm B}$ when $T_{\rm HII}=6\times 10^4\,$K and $T_\infty \sim 3\times 10^2\,$K due to dust cooling. For comparison, the mean accretion rate for the case of pristine gas is $\sim 1\%$ of the Bondi rate because the temperature ratio is $\delta T \sim 6$ with $T_\infty \sim 10^4\,$K \citep{ParkR:12}.

\begin{figure*}
   \epsscale{1.16}
   \plottwo{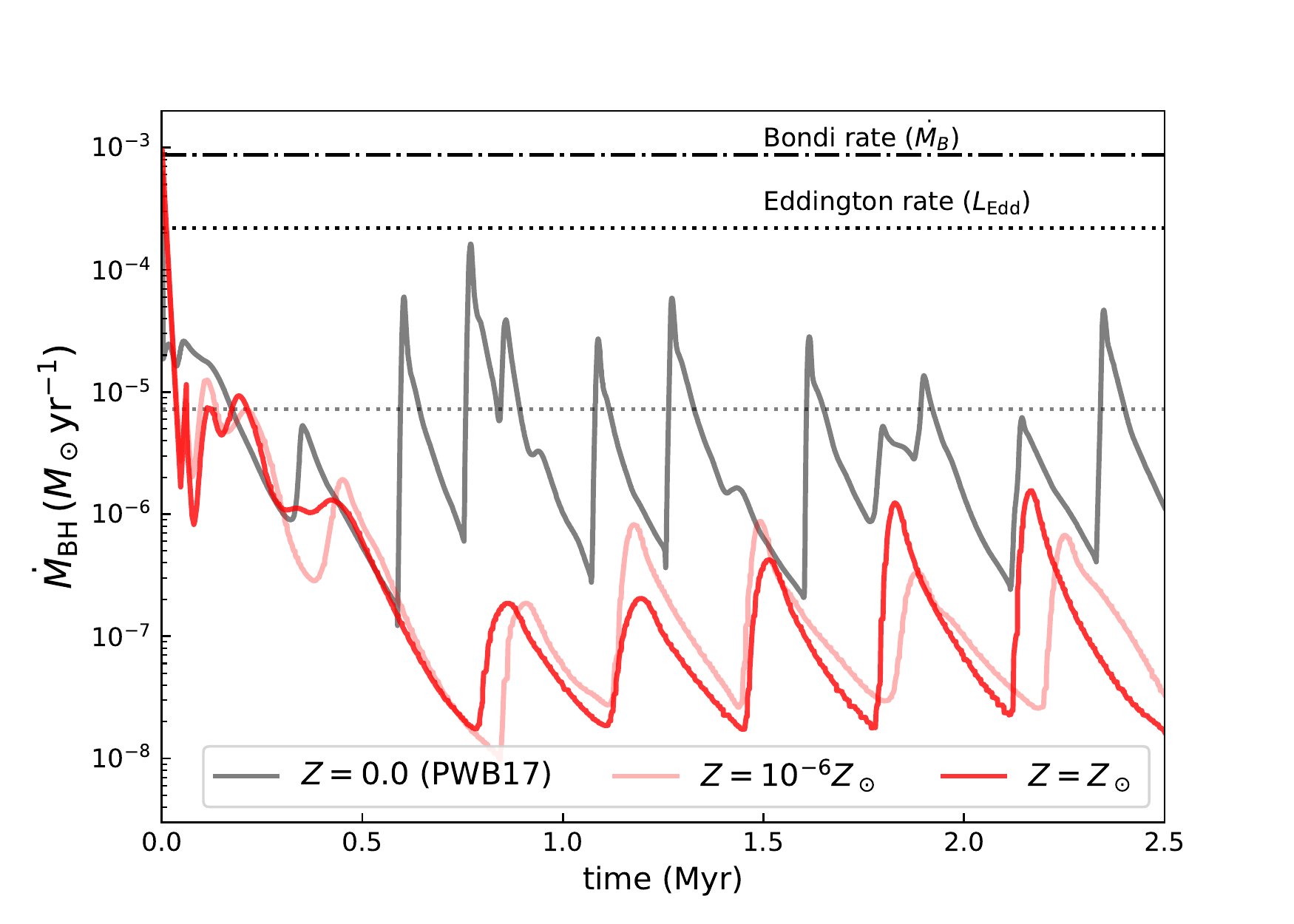}{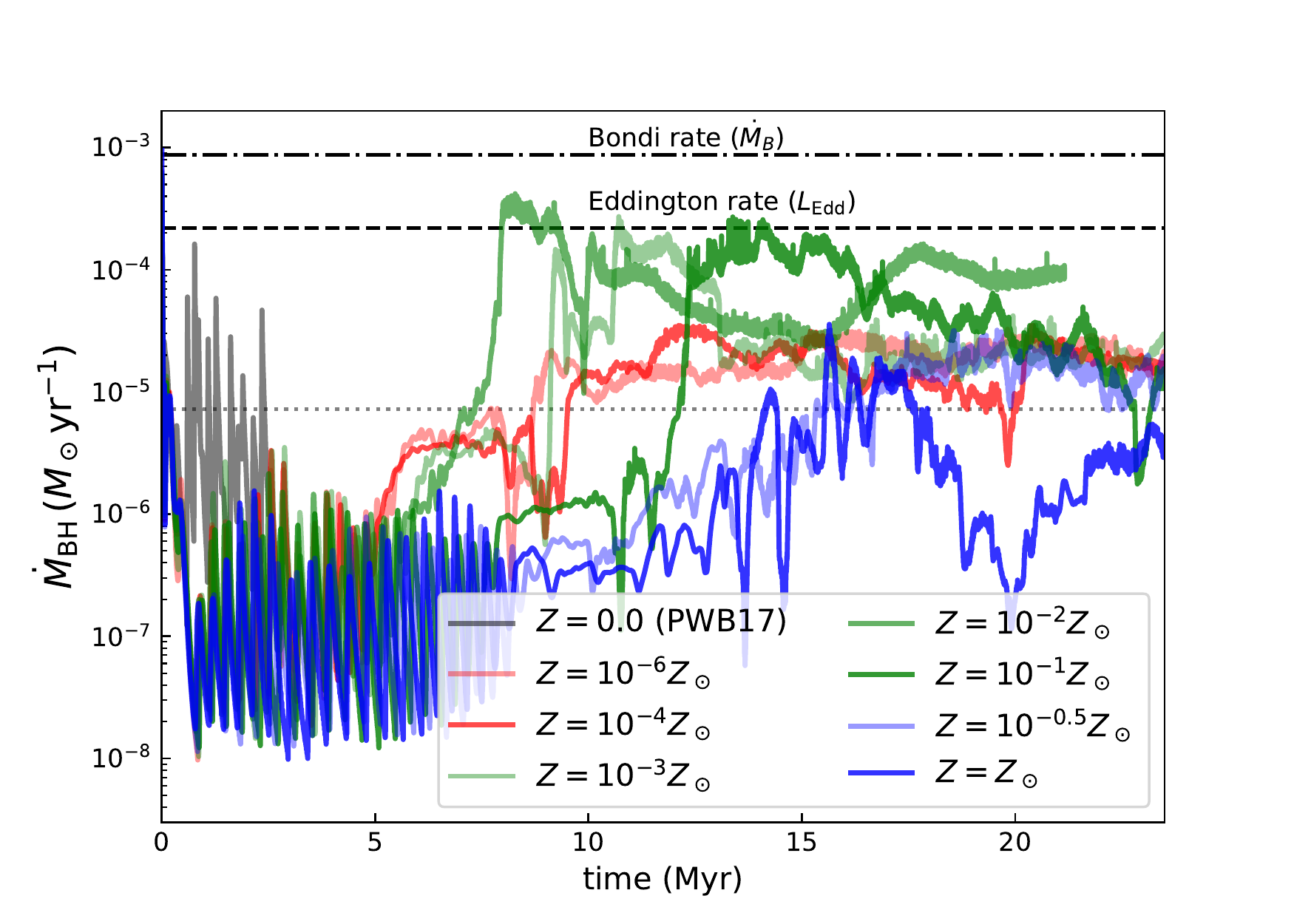}
      \caption{
      Left: accretion rate as a function of time for Zm0 (dark red) and Zm6 (light red).  The gray line is from the simulation with zero metallicity in PWB17 and the average rate is shown as a dotted line. The accretion rate of the early phase is characterized by the lower averages compared to zero metalliciy case and the oscillatory behavior due to the cycle between accretion and feedback for all the range of metallicities. Right: the late phase of the simulations at $t \gtrsim 8\,$Myr is characterized by stable and enhanced accretion rates.} 
      \label{fig:accrate}
\end{figure*}


\subsection{Size of the ionized region}

As the BH accretes and shines, an $\hii$ region forms, and in this paper we track how it evolves.  Below we describe the connection between the BH accretion luminosity and the size of the resulting $\hii$ region. The BH luminosity is calculated from the accretion rate as
\begin{equation}
   L_{\rm BH}= {\rm min} \left(\eta \dot{M}_{\rm BH} c^2, L_{\rm Edd}\right),
\end{equation}   
where we adopt the radiative efficiency of $\eta=0.1$ assuming a thin disk model \citep{ShakuraS:73} and $c$ is the speed of light. We simply adopt the Eddington luminosity as the cap for the luminosity since the accretion rates mostly stay below $L_{\rm Edd}$. However, it is worth noting that global 3D
magneto-hydrodynamic simulations show that the accretion luminosity converges to in the order of $10\,L_{\rm Edd}$ even when $\dot{M}_{\rm BH} \gtrsim 10^3\,L_{\rm Edd}/c^2$ \citep{Sakurai:2016,Jiang:2019}. The Eddington luminosity is
\begin{equation}
   L_{\rm Edd} = \frac{4\pi G\mbh m_{\rm p}c}{\sigma_{\rm T}} \simeq
   1.26\!\times\!10^{38}
   \left(\frac{\mbh}{\msun}
   \right)\,{\rm erg\ s}^{-1},
\end{equation}   
where $m_{\rm p}$ is the proton mass and $\sigma_{\rm T}$ is the Thomson cross-section for electrons. Therefore the mean size of an ionized sphere is calculated as
\begin{equation}
   \ars =
   \left( \frac{3 N_{\rm ion}}{4\pi \alpha_{\rm rec} n_e n_{\rm H}}  \right)^{1/3},
   \label{eq:Rs}
\end{equation}
where $N_{\rm ion}$ is the total number of ionizing photons in the range of $13.
6\,$ eV $\le E \le 100$\,keV with a spectral index of $\alpha_{\rm spec}=1.5$ 
for power-law energy distribution, and $\alpha_{\rm rec}$ is the case B 
recombination coefficient. In the case of pristine gas, comparing Equations~(\ref
{eq:bondiradius}) and (\ref{eq:Rs}), the parameter space that we explore with 
$\mbh = 10^4\,\msun$ and $\nH = 10^3\,{\rm cm}^{-3}$ belongs to the 
feedback-limited regime as $\mbh \nH = 10^7\,\msun\,\cmq$ that is less than the critical threshold of $10^9\,\msun\, \cmq$ \citep
{ParkRDR:14a}.

Photo-ionization, photo-heating, and gas cooling are computed by the adaptive 
ray-tracing module {\it Moray} that couples these rates to the hydrodynamic 
equations (see section 2.2 of PWB17). Here, we use 4 energy bins of (28.4, 263.0, 2435.3, 22551.1) eV, each with a fractional luminosity of (0.6793, 0.2232, 0.0734, 0.0241), respectively (see section 2.3 in PWB17 for details).

\subsection{Dust modeling}
\label{sec:dust_modeling}
For a given metallicity $Z$, the initial carbon and iron abundances are set as
$A({\rm C})=8.43$ and $A({\rm Fe})=7.50$ \citep{Asplund:2009} for the solar metallicity, where we use $Z_\odot=0.02$ \citep{Grevesse:1998}.\footnote{The abundance of an element $X$ is defined as $A(X) = 12 + \log(y_X/y_{X, \odot})$, where $y_X$ is the number fraction relative to hydrogen nuclei. Note that the metallicity $Z_\odot = 0.01295$ \citep{Asplund:2009} would be more consistent with the elemental abundances that we use.}

For compact, spherical dust grains with a radius $a_{\rm d}$, we calculate the dust temperature $T_{\rm d}$ from the thermal balance between radiation heating, collisional heat transfer from gas to dust and dust thermal emission as
\begin{equation}
\frac{\pi a_{\rm d}^2}{m_{\rm d}} \left[ \int  F_{\nu} Q_{\nu} d\nu + v_{\rm th} n_{\rm H} (2kT - 2kT_{\rm d}) \right] = 4\sigma_{\rm SB} \kappa_{\rm P,d} T_{\rm d}^4,
\end{equation}
where $F_{\nu}$ is the flux of radiation from a BH. 
$Q_\nu$ is the absorption coefficient, and
$\kappa_{\rm P,d}$ is the Planck-mean opacity of dust per unit dust mass taken from \citet{Nozawa:2008}. $m_{\rm d}= \frac{4\pi}{3} \varrho_{\rm d} a_{\rm d}^3$ is the mass of a dust particle, where $\varrho_{\rm d}$ is the bulk density of a grain. We use $\varrho_{\rm d} = 3.20$ and $2.28~{\rm g~cm^{-3}}$ for silicate and graphite, respectively.
The gas cooling rate, H$_2$ formation rate and grain opacity are calculated with the derived dust temperature \citep[in more detail, see][]{Chiaki:2015}. Silicate and graphite sublimate above the dust temperatures $1222\,$K and $1800\,$K, respectively \citep{Pollack:1994}. We model that nuclei tied up in grains are released into the gas phase as
\[
{\rm MgSiO_3 (s)} + 2{\rm H_2} \to {\rm Mg} + {\rm SiO} + 2{\rm H_2O},
\]
\[
{\rm C (s)} \to {\rm C},
\]
where ``s'' denotes the solid phase.

We add the radiation pressure on dust grains of 
uniform size in addition to the radiation forces on free electrons and neutral hydrogen. The optical depth is expressed as
\begin{equation}
 d \tau_{d, \nu} = Q_\nu \pi a_{\rm d}^2 \frac{D m_{\rm H} \nH}{m_{\rm d}} dl = 
 \frac{3Q_\nu m_{\rm H} \nH}{4\rho_{\rm d} a_{\rm d}} dl , 
\end{equation}
where 
$m_{\rm H}$ is hydrogen mass, $\rho_
{\rm d}$ is the dust mass density, and $D$ is the dust-to-gas mass ratio. We 
use the value $D \equiv M_{\rm dust}/M_{\rm H}$ from each cell. $D$ is set to $0.005 (Z/Z_{\odot})$ for both silicate and graphite.
We assume a single-sized grain model with $a_{\rm d} =0.1\,\mu$m. 
Note that observations of dust in the Milky Way indicates 
that the size distribution is well approximated by the power-law \citep{MRN:1977}
\begin{equation}
\frac{d n_{\rm d}}{d a_{\rm d}} \propto a_{\rm d}^{-3.5}.
\end{equation}

The radiation pressure on dust is much larger than the case for Compton scattering on electrons by a factor of
\begin{equation}
   \frac{f_{\rm rad}^{\rm dust}}{f_{\rm rad}^{\rm e}} 
   \sim 7.1\times 10^{2} \left(\frac{a_{\rm d}}{\rm 0.1\,\mu m} \right) ^{-1} \frac{Z}{Z_\odot}, 
\end{equation}
which is proportional to the metallicity $Z/Z_\odot$ \citep{Yajima:2017}. This ratio indicates that the radiation pressure on dust dominates over the case for electron for $Z \gtrsim 10^{-3}\,Z_\odot$. UV photons are absorbed and re-emitted in the IR by dust grains, which can then escape from the system without much absorption. In this study, we do not 
consider the radiation pressure by the IR, however, it should be considered 
when the column density is above $10^{22}\,{\rm cm}^{-2}$.

\section{Results}
\label{sec:results}

All the simulations listed in the Table.~\ref{table:para} with different 
metallicities which span 6 orders of magnitude, show a similar evolution 
pattern. They start from a highly {\it oscillatory quasi-spherical mode} that
we refer to as the {\it early phase} and make a transition to {\it high accretion mode with broken symmetry} with smaller variations that we refer to as the {\it late phase}.

\subsection{The Early Phase}
\subsubsection{Accretion rates evolution}

The left panel of Fig.~\ref{fig:accrate} shows the early evolution of accretion rates compared to the previous simulations with zero-metallicity (grey line) from PWB17. The light red line is for Zm6 with $Z=10^{-6}Z_\sun$ while dark red is for Zm0 with $Z=Z_\sun$. Bondi accretion rate for the temperature of $T_\infty=8\times 10^3\,$K (dot-dashed) and Eddington rate for the $\mbh=10^4\,\msun$ with $\eta=0.1$ (dashed) are shown for references. All the simulations immediately start from decreasing accretion rates and the runs with dust drop to a much lower minimum rate of $\dot{M}_{\rm BH}\sim 10^{-8}\,\dot{\rm M}_\sun {\rm yr}^{-1}$ at $t \sim 0.8\,$Myr. The accretion rates shows a repeated pattern of fast rising and exponential decaying since then.

\begin{figure*}
   \includegraphics[width=1.0\linewidth]{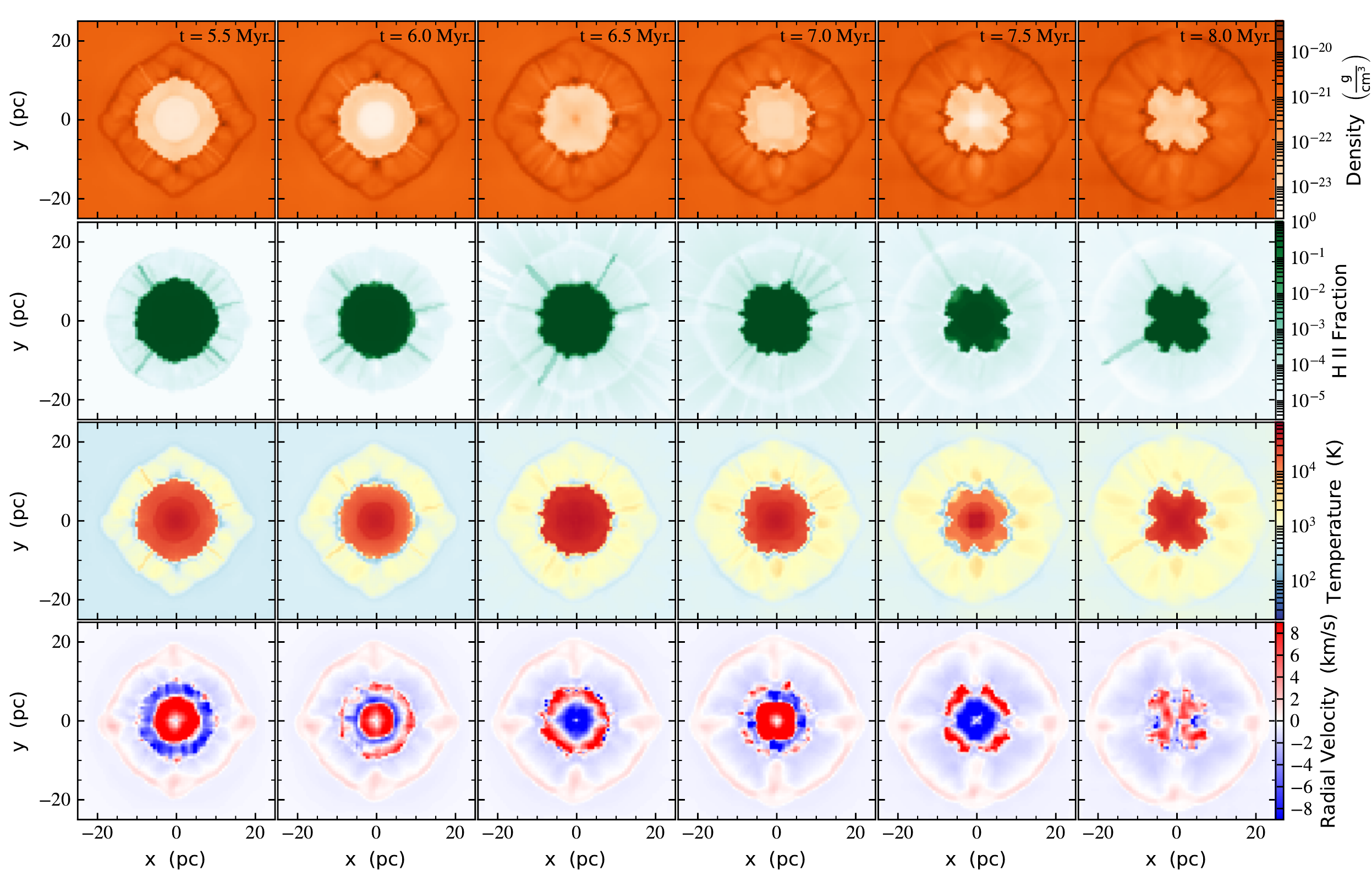}
   \caption{Early evolution for Zm0 run. From top to bottom, the panels show slices of density, $\hii$ fraction, temperature, and radial velocity at $t=5.5$, 6.0, 6.5, 7.0, 7.5, and 8.0\,Myr (left to right). The early phase is characterized by the oscillatory behavior of the ionized region due to the accretion and feedback cycles. 
   }
   \label{fig:evol_early}
\end{figure*}

Previous simulations with zero-metallicity find that the mean accretion rate is approximately 2 orders of magnitude lower than the classical Bondi accretion rate for $T_\infty=10^4\,$K. With dust included in the study, the mean accretion rates are approximately two orders of magnitude lower than the case for zero-metallicity gas. Note that the mean accretion rate for pristine gas is $\sim 7\times 10^{-6}\,\msun\,{\rm yr}^{-1}$, whereas with dust it is $\sim 2\times 10^{-7}\,\msun\,{\rm yr}^{-1}$ as shown in the 4th column of Table\,\ref{table:para}. Using Eq.~(\ref{eq:mdot}) to compare the current accretion rate relative to the zero-metallicity case, we find that it scales to
\begin{equation}
 \langle \dot{M}_{\rm BH, d} \rangle= \langle \dot{M}_{\rm BH, 8000K} \rangle \left( \frac{8000\,{\rm K}}{300\,{\rm K}} \right)^{3/2} 
 \left(\frac{\delta T_{\rm d}}{\delta T}\right)^{-5/2}, 
 \label{eq:mdotdust}
\end{equation}
where $\dot{M}_{\rm BH, 8000K}$ is the accretion rate for $T_\infty=8000\,$K. With the temperature contrast between the ionized and neutral region $\delta T \sim 6$ and $\delta T_{\rm d}\sim 200$, the ratio becomes $\langle \dot{M}_{\rm BH} \rangle / \langle \dot{M}_{\rm BH, 8000K} \rangle \simeq 0.02$ which is consistent with the mean accretion rates in the simulations. 

\subsubsection{Oscillatory behavior in the early phase}
Fig.~\ref{fig:accrate} shows that the period of bursting accretion is similar to the case of the pristine gas despite the much lower accretion rates. This similarity is explained by the pressure equilibrium across the ionization front. It is affected by the change in temperature in the following quantities in Eq.~(\ref{eq:mdotdust}) that originate from the Str{\"o}mgren radius Eq.~(\ref{eq:Rs}): the electron number density $n_e=(\delta T/\delta T_{\rm d}) n_{\rm H, \infty}$ (considering a completely ionized medium), the hydrogen number density $n_{\rm H}=(\delta T/\delta T_{\rm d}) n_{\rm H, \infty}$, the recombination rate $\alpha_{\rm rec} \propto T_\infty^{-1/2}$, and the ionizing photon luminosity $N_{\rm ion}=0.02 N_{\rm ion, 8000}$. The last relation comes from the decreased accretion rate as discussed in the previous subsection. The mean size of the dust-affected ionized region is
\begin{equation}
   \ars_{\rm d}
   =  0.02^{\frac{1}{3}}\left(\frac{\delta T_{\rm d}}{\delta T}\right)^{\frac{2}{3}} \left(\frac{300\,K}{8000\,K}\right)^{\frac{1}{6}} \ars_{8000} \simeq 1.6, 
\end{equation}
which expect 60\% larger in radius compared to PWB17. This size is consistent with PWB17 when compared to the average size of $\ars$. The behavior is also consistent in terms of the average period between the cycles as Zm0 and Zm6 runs show longer periods in the left panel of Fig.~\ref{fig:accrate}. This is due to the fact that the period is proportional to the $\ars$ assuming a similar gas depletion rate inside the \hii~region \citep[see][for details]{ParkR:11,ParkR:12}.

\begin{figure*}
   \includegraphics[width=1.0\linewidth]{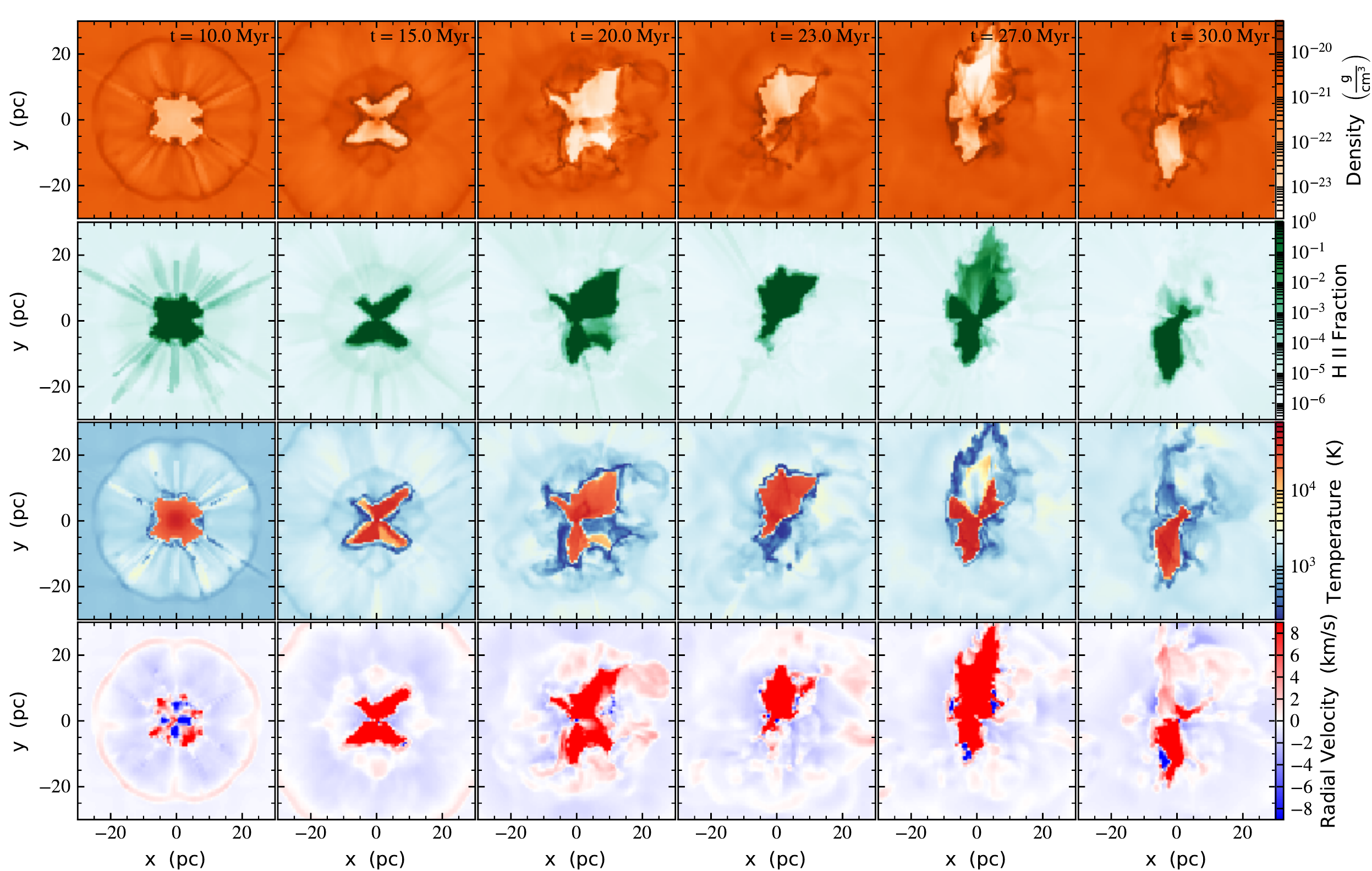}
   \caption{Late phase evolution for Zm0. Similar to Fig.~\ref{fig:evol_early}, but at late times $t=10.0$, 15.0, 20.0, 23.0, 27.0, and 30.0\,Myr (left to right). The late phase is characterized by the broken quasi-spherical symmetry of the ionized region with strong outflows and cold dense gas structure with random shapes.}
   \label{fig:evol_late}
\end{figure*}

\subsubsection{Evolution of thermal structure}
Fig.~\ref{fig:evol_early} shows the density, $\hii$ fraction, temperature, and radial velocity snapshots at $t=5.5$, 6.0, 6.5, 7.0, 7.5, and 8.0\,Myr. Ionizing radiation from the BH creates a low-density, highly ionized, and hot bubble, similar to the cases of PWB17. However, the temperature structures show a clear difference. At scales larger than 20\,pc, the gas temperature is not affected by the radiation, and a spherical density front forms from a fluctuation in the ionized region. The gas temperature drops to $T_\infty \sim 3\times 10^{2}$\,K from the initial condition of $T_\infty =10^4$\,K. The lower gas temperature and associated lower thermal pressure decrease the mean accretion rates \citep{ParkR:12}. The radial velocity snapshots in the bottom row clearly show the fluctuation of the ionized region due to the cycles between accretion and feedback. The fluctuation drives density waves which are seen as enhanced density shells at $r \sim 20$\,pc. Also, note that high-energy X-ray photons selectively propagate out of ionized region while most of the low-energy UV photons are trapped inside the low-density region. Streaks of enhanced $\hii$ fraction are observed in the $\hii$ fraction snapshots at $10 < r < 20$\,pc. This creates the intermediate temperature layer of $T_{\rm int} \sim 10^3$\,K between the gas reservoir and the ionized region.

There is another thin layer of low temperature gas just outside the ionized region. Gas inflow from reservoir and outflow due to radiation feedback converge just outside the $\hii$ region and increase the gas density. Enhanced gas density promotes the radiative gas cooling further.

\subsection{Late Phase}
\subsubsection{Stable and enhanced accretion rates}
The right side of Fig.~\ref{fig:accrate} shows the early and late phase accretion
rates up to $t=23.5$\,Myr. The early phase of the accretion rates of all the 
simulations with different metallicities is characterized by the oscillatory 
behavior while the late phase accretion rates are increased continual accretion 
rates. The accretion pattern makes a noticeable change at $t \sim 7.2$\,Myr for 
Zm0 while the transition of the other runs occurs earlier around $t\sim 4.
7$\,Myr. All of the transition times $\tau_{\rm tran}$ are listed in the 3rd column of Table~\ref{table:para}, generally occurring later for higher metallicities. The oscillatory accretion rates in the early phase vary less and the overall accretion rates increase. The run Zm0 shows a longer variation period, e.g., the accretion rate is $\dot{M}_{\rm BH}\sim 3\times 10^{-5}\,{\rm \msun}\,{\rm yr}^{-1}$ at $t\sim 17$\,Myr and drops to $\dot{M}_{\rm BH}\sim 10^{-7}\,{\rm \msun}\,{\rm yr}^{-1}$ at $t\sim 20$\,Myr. Other runs also show the minor variation of accretion rates, but clearly show a higher average than Zm0. Last column of Table~\ref{table:para} shows the mean accretion rates $\langle 
\dot{M}_{\rm late} \rangle$ for $10.0 \le t \le 25.0\,$Myr. Note that the 
average accretion rates are $\ge 10^{-5}\,\msun\,{\rm yr}^
{-1}$ which is higher than the $\langle 
\dot{M}_{\rm early} \rangle$ by $\sim 2$ orders of magnitude except for Zm0.

\subsubsection{Asymmetric thermal structure}

\begin{figure*}
   \epsscale{1.15}
   \plottwo{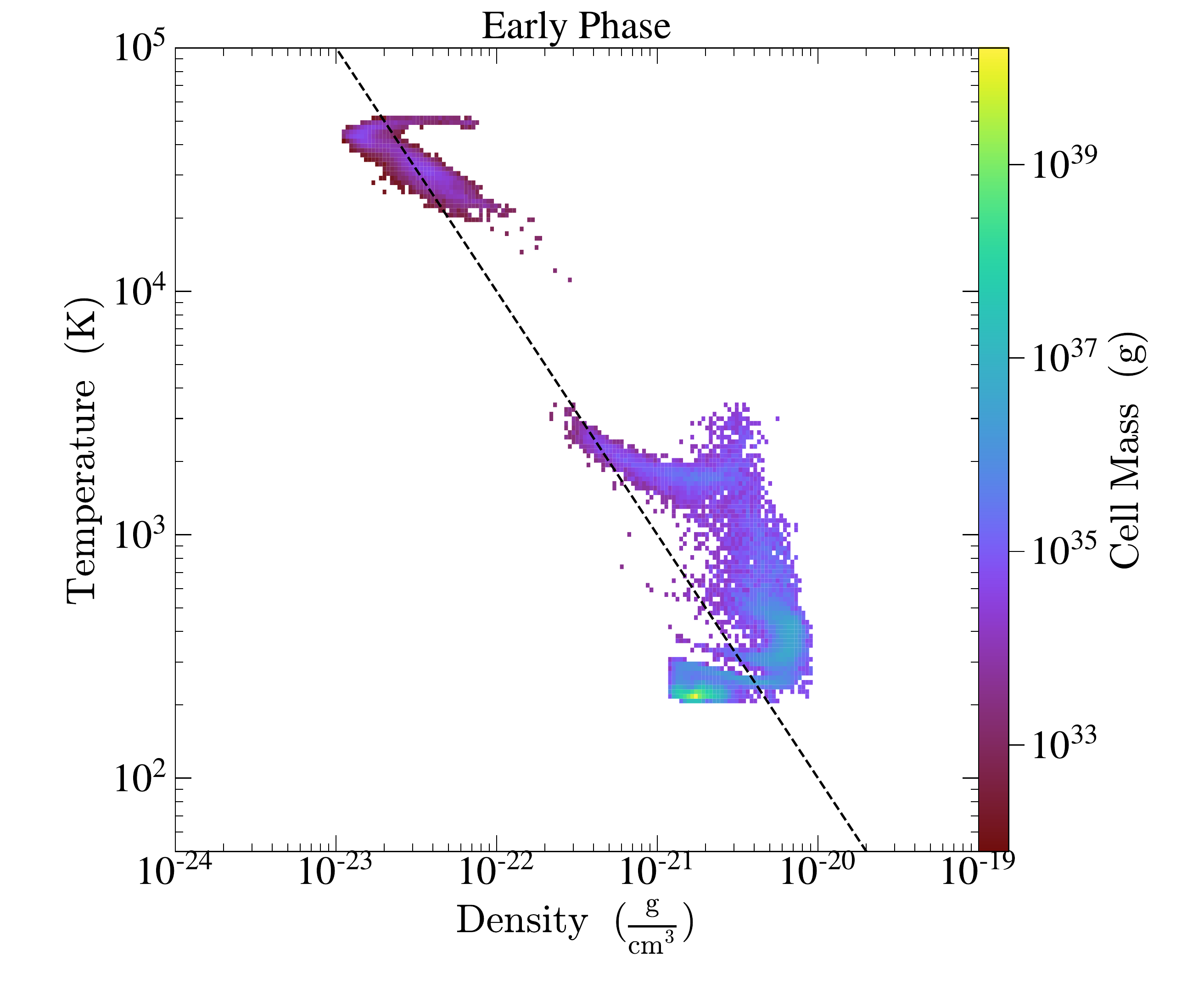}{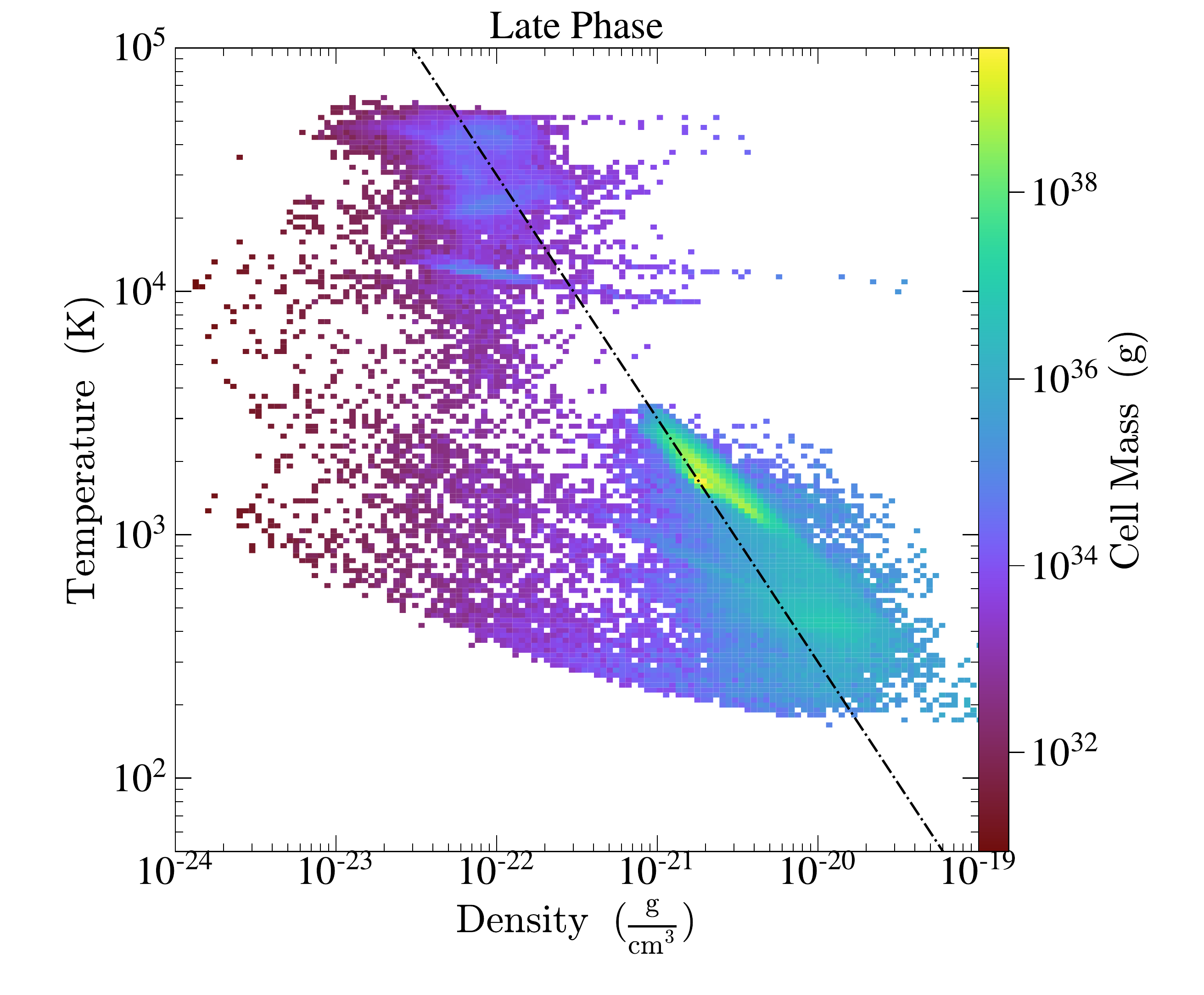}
   \caption{Phase diagrams for early (left) and late (right) phases for Zm0 at $t=2.2$ and $17.0$\,Myr, respectively. The pressure equilibrium lines from Eq.~(\ref{eq:pressure_early}) and (\ref{eq:pressure_late}) are shown for each phase.}
   \label{fig:phase}
\end{figure*} 

Fig.~\ref{fig:evol_late} shows the snapshots in the late phase at $t=10.0$, 15.0, 
20.0, 23.0, 27.0, and 30.0\,Myr. The snapshot at $t=10.0$\,Myr shows the moment 
when the transition to the late phase starts. Here the thin dense 
shell of gas becomes clumpy, allowing for the X-ray photons to leak preferentially through the porous medium, which heat and ionize the affected regions.

The radial velocity slices show that inflows and outflows develop in 
different directions. The small outflow structure grows in size at $t=15.0\,$Myr. The increasing BH luminosity can not drive the gas clumps away from the 
central BH. In the subsequent snapshots, the strong outflow and high-density inflow grow simultaneously. The ionized region contains strong 
outflows that are surrounded by a cold dense shell. The morphology of the ionized region at $t \ge 20\,$Myr becomes highly random 
depending on the direction of the outflows.

\subsection{Evolution of thermal pressure}

Fig.~\ref{fig:phase} shows the phase diagrams in the early ($t=0.22\,$Myr) and late phase ($t=17.0\,$Myr) for the gas within $r \le 40$\,pc for tor Zm0 run. In the early phase, most of the gas cools to $T\sim 3\times 10^{2}$\,K, and the temperature of the gas inside the ionized region increases to $T=2\,$--$\,6 \times 10^{4}$\,K. In the early phase, the dashed line shows the line for the pressure equilibrium 
\begin{equation}
{p}_{\rm early} \propto \rho_{\rm gas} T_{\rm gas} \sim 1 \times 10^{-18} {\rm g}\,{\rm cm}^{-3}\,{\rm K}\,.
\label{eq:pressure_early}
\end{equation}
Pressure equilibrium is maintained between the neutral and ionized region, and the gas above the line is the dense gas located outside the ionized region shown in Fig.~\ref{fig:evol_early}. In the late phase, the gas temperature of the neutral region increases to $T \sim 2\times 10^3$\,K meaning that the thermal pressure also increases by $\sim 1$ order of magnitude compared to the early phase. The dot-dashed line shown in the right phase diagram is
\begin{equation}
   {p}_{\rm late} \propto \rho_{\rm gas} T_{\rm gas} \sim  3 \times 10^{-18} {\rm g}\,{\rm cm}^{-3}\,{\rm K}\,.
\label{eq:pressure_late}
\end{equation}
The gas below the thermal equilibrium line is the transitional outflow which is supported by the momentum in addition to the thermal pressure. Most of the gas is generally in pressure equilibrium, however, momentum-driven outflows drive more gas above this pressure equilibrium line.

\subsection{Thermal structure of gas and dust}

\begin{figure*}
   \begin{center}
   \includegraphics[width=0.6\linewidth]{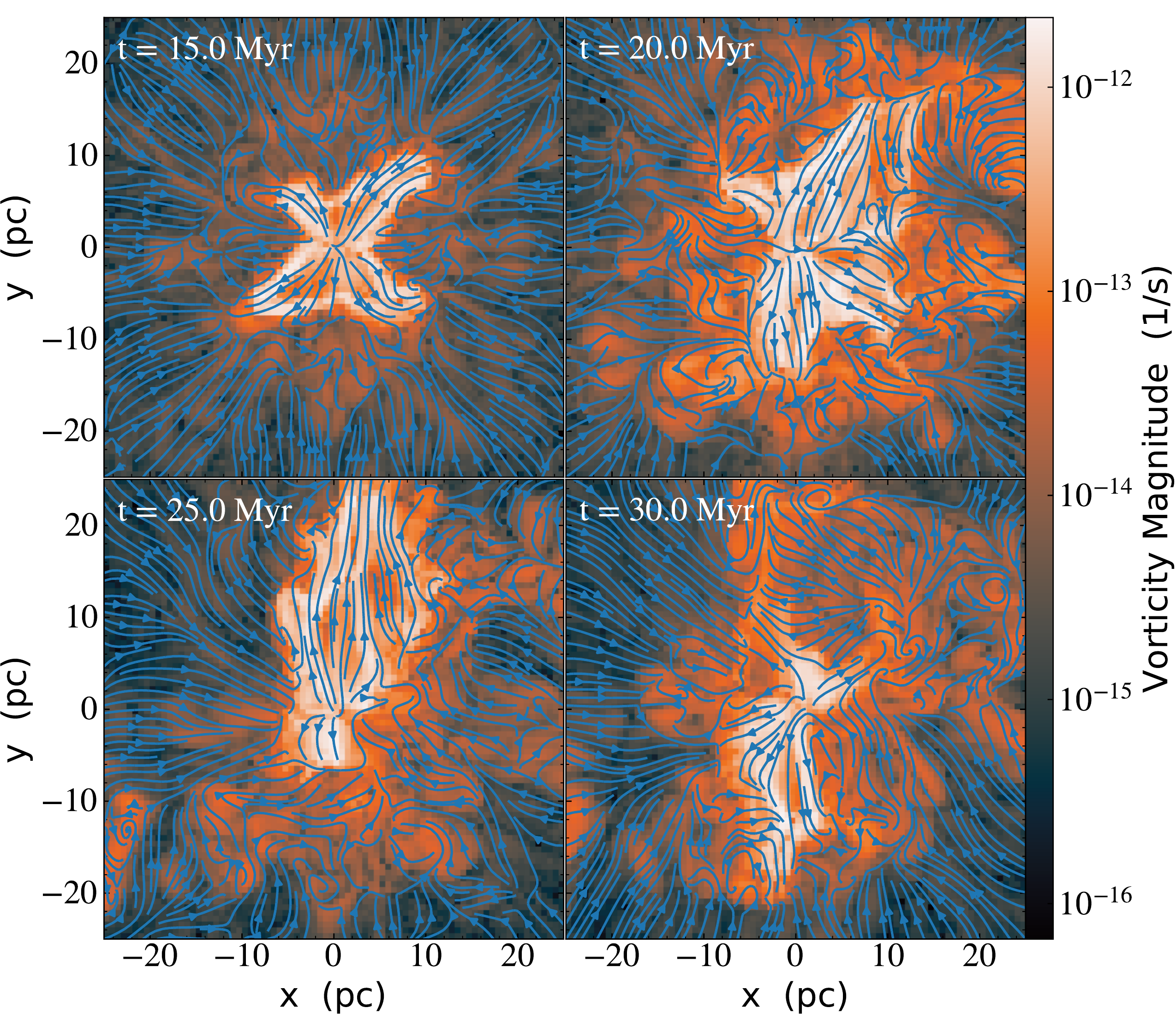}
   \caption{Slices of vorticity magnitude $|\vec{\omega}|$ with overlaid velocity streamlines at times $t=15$, 20, 25, and 30~Myr for Zm0.
   Strong outflows within the ionized region and inflows from a larger scale converge at the ionization front and create eddies.} 
   \label{fig:vel_stream}
   \end{center}
\end{figure*}

The vorticity $\vec{\omega} = \vec{\nabla} \times \vec{v}$ provides a measure of turbulence in units of ${\rm s}^{-1}$ that is approximately the inverse eddy turnover time. Fig.~\ref{fig:vel_stream} shows the magnitude of vorticity $|\vec{\omega}|$ with velocity streams at $t=15$, 20, 25, and 30\,Myr for Zm0 run. Outflows dominate the interior of the ionized region while low-velocity inflows are dominant at larger scales. Just outside of the ionization front, the outflows and inflows converge, driving turbulence and causing the outflow structure and direction are highly variable. 

\begin{figure*}
   \includegraphics[width=\linewidth]{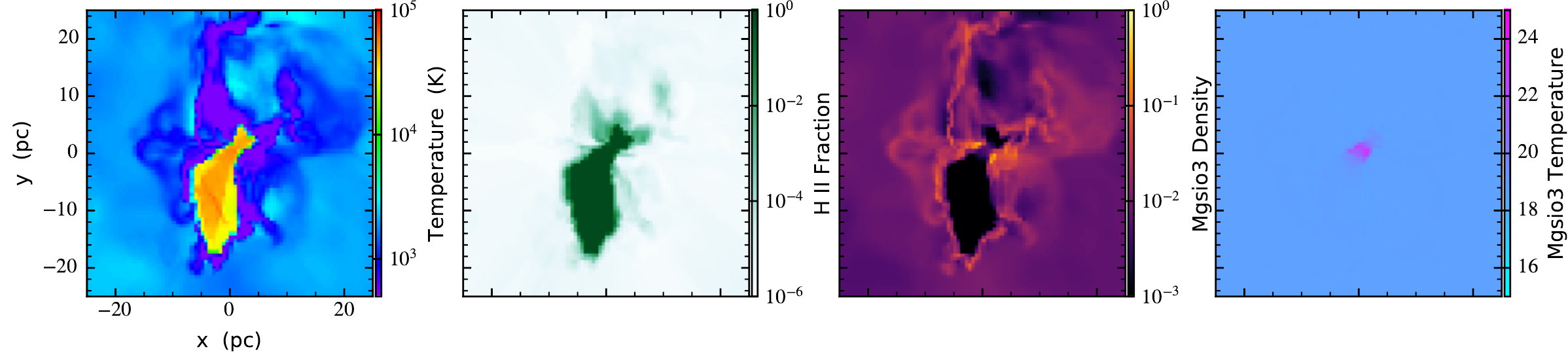}
   \caption{From left to right, panels shows slices of gas density, $\hii$ fraction, and $\enstatite$ density and temperature for Zm0 at $t=30$\,Myr. }
   \label{fig:last}
\end{figure*}

Fig.~\ref{fig:last} shows slices of the gas temperature, $\hii$ fraction, and $\enstatite$ density and temperature from left to right at $t=30$\,Myr for Zm0 run. The thermal structure of the ionized region is highly asymmetric depending on the stage of the evolution. As discussed earlier, the gas is characterized by three different temperature structures. The ionized region has the highest temperature at $T_{\rm HII}\sim 6\times 10^4$\,K that is surrounded by cooler dense gas at $T\sim 10^3$\,K. The \hii~fraction shows that the high-density gas is partially ionized by the leaking X-ray photons. The $\enstatite$ density, shown as a fraction of the initial gas density, has a similar structure to the gas density because we use a fixed ratio of dust-to-gas. The dust density is depleted inside the ionized region with an average dust temperature of $T_{\enstatite} \sim 20\,$K and a minor enhancement only nearby the BH. The dust temperature increases only in the high-density region and not in the ionized region because of its low density. Dust sublimation is not captured by our simulations because we do not resolve the sublimated regions that have $T_{\rm d} > 1{,}500\,$K.

\begin{figure}
   \includegraphics[width=1.05\linewidth]{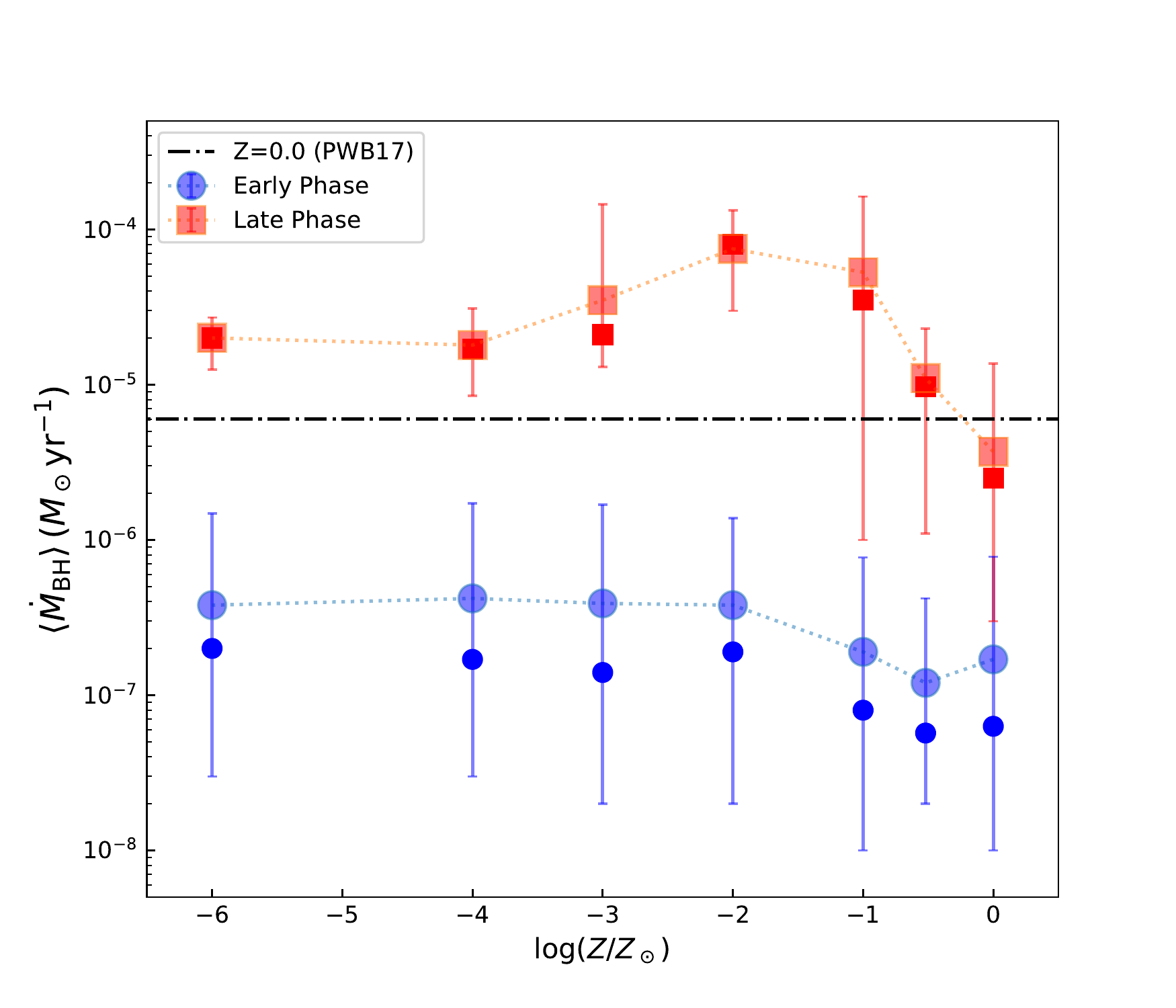}
   \caption{Mean accretion rates for the early (blue circles) and late phase (red squares). Small circles and squares show the medians for each phase, respectively. Error bars show the 5\%/95\% (2$\sigma$) percentiles. Dot-dashed line is from the simulation with zero metallicity from PWB17.}
   \label{fig:mean_accrate}
\end{figure}

\subsection{Mean accretion rates as a function of metallicity}
Fig.~\ref{fig:mean_accrate} shows the mean accretion rate in physical units as 
a function of metallicity. Blue circles show the accretion rates at early 
phase $\langle \dot{M}_{\rm early} \rangle$ while red squares show the mean 
rate at late phase $\langle \dot{M}_{\rm late} \rangle$. The values of $\langle \dot{M}_{\rm early} \rangle$ for each metallicity is calculated by taking the mean rate at $2.0 \le t \le 4.0\,$Myr while the mean at $10.0 \le t \le 25.0\,$Myr is taken for $\langle \dot{M}_{\rm late} \rangle$. The $\langle \dot{M}_{\rm early} \rangle$ and $\langle \dot{M}_{\rm late} \rangle$ are listed in the 4th and 5th columns of Table~\ref{table:para}. The upper and lower limits for the mean rates shown as error bars represent the 5\%/95\% (2$\sigma$) percentiles, respectively. The small symbols show the medians for each phase and the offset from the mean value indicates that the distribution is not Gaussian.

In the early phase, the accretion rates are about an order of magnitude lower 
than the case of zero metallicity, shown as a dot-dashed 
line. It becomes even lower with high metallicity runs with $Z \ge 10^{-1}
Z_\odot$, consistent with \citet{Yajima:2017}, who find that
the mean accretion rate is one order of magnitude lower than the case for 
primordial gas for $Z \le 0.01Z_\sun$. The standard deviations for each 
metallicity are shown as error bars which do not vary much as a function of 
metallicity. 

In the late phase, the mean accretion rates $\langle \dot{M}_{\rm late} \rangle$ are in general higher than the zero-metallicity case, and it is the highest for $Z=10^{-2}Z_\sun$ by about an order of magnitude. $\langle \dot{M}_{\rm late} \rangle$ increases with metallicity for low metallicities until it peaks $Z \sim 10^{-2}Z_\odot$. The $Z \sim 10^{-1}Z_\odot$ case has similarly high rates, however, the mean accretion rate drops below the primordial case when the metallicity reaches the solar value $Z=Z_\odot$. 

\section{Discussion}
\label{sec:discussion}
We have built upon the previous work of \citet{Yajima:2017}, finding that the early phases of our simulations are in good agreement. By evolving the simulations significantly longer, we find that the accretion transitions to an increased and less variable rate. This late phase might last until the available gas is consumed by star formation or BH accretion. With increasing mass, the BHs will become more susceptible to hyperaccretion.

Cosmological simulations with star formation and feedback, the inclusion of dust, and accretion and feedback from seed BHs will provide insightful clues toward a more complete picture of BH accretion lifecycles in the early universe. Our simulations are idealized to focus on the dust physics on the growth of seed BHs, however, the dust and metals produced from supernovae and the radiative feedback from the BHs should be considered together to understand the role of dust contamination on BH growth. In particular, we found that in the early phases of accretion, pressure equilibrium across the ionization front regulates the gas accretion rates, and accordingly, the accretion rate estimate given in Eq.~(\ref{eq:mdotdust}) matches with the simulations. However, in the late phase, the spherical symmetry is broken, and thus the accretion rates from Eq.~(\ref{eq:pressureequilibrium}) only apply to the early phase.

Furthermore, our simulations are capable of tracking the sublimation of the dust for graphite and silicates, however, the sublimation of dust might occur only at the small distance from the BHs, which is not resolved in our simulations. Dust temperatures increase near the BH, but at the levels of $T_{\rm d}\sim 30$~K. Dust sublimation near the BHs seems to have a minor impact on the accretion rate due to the low fraction of the dust compared to hydrogen and helium gas. Thus, the accretion rates from our simulations should not be affected greatly in higher resolution simulations that better captures dust sublimation.

Our simulations include physical processes over a range of temperatures and photon energies, such as a non-equilibrium chemical network with dust species and the radiation transport of X-rays.  The resulting physical structures will therefore provide solid basis for observational predictions.  Existing radio telescopes, such as ALMA, and space-based IR telescopes, such as JWST, may detect the signatures of the rapidly growing BHs in the early universe. For example, \citet{Yajima:2017} describe an IR observational signature, the flux ratio of $\lesssim 20\,\mu m$ and $\gtrsim 100\,\mu$m, that is closely related to the Eddington ratio.  Radiative transfer calculations \citep[e.g.][]{Narayanan:2021} compute detailed SEDs from the stars, black holes, and gas, ranging from X-rays to the sub-mm, using the simulation physical fields as input.  Our simulated systems are highly variable and turbulent and have anisotropic features in the temperatures and abundances of different dust species and will vary to some degree depending on viewing angle. The X-ray photons that leak through the porous medium could be detectable by future X-ray missions, such as ATHENA, will also provide clues about actively accreting seed BHs.  We will address such observational signatures in a later study.

\section{Summary}
\label{sec:summary}
In this study, we run a suite of 3D RHD simulations using {\it Enzo}+{\it Moray} with dust physics to study the role of dust in enhancing the growth of seed BHs in the early universe. We find that a dusty ambient medium, compared to the zero-metallicity case, induces changes in the thermal structures surrounding the BH, leading to a change in accretion behavior. Our simulations show distinctive phases as a function of time (i.e., the early and late phases). These two phases are different in the accretion rates and thermal structure of the neighboring gas. The following list the main discoveries.

\begin{itemize}
\item {\bf The early phase}: Radiative cooling by dust decreases the gas temperature below $T\sim 3\times 10^2$\,K, and the associated drop in thermal pressure reduces the mean accretion rate by approximately two orders of magnitude compared to the case of chemically pristine gas.  Nevertheless, radiation from the accreting BH ionizes and heats the surrounding gas. The system cycles between accretion and feedback modes, causing oscillatory behavior in both the accretion rates and $\hii$ region size.

\item On average, thermal pressure equilibrium across the ionization front is maintained, and the density in the \hii~region is low due to the temperature contrast across the ionization front. Low densities and high recombination rates make the size of the ionized region comparable to the pristine case despite the low accretion rate. 

\item Outflows within the ionized region and inflows from a larger scale converge at the ionization front and enhance the gas density. Hard X-rays leak into the neutral region, heating and ionizing the gas, that result in distinctive regions with varying temperatures.

\item {\bf Transition}: The high-density gas outside the ionization front becomes clumpy and develops outflow and inflow channels. As the high-density gas fuels the BHs, the accretion rate and thus the BH luminosity increase rapidly.

\item {\bf The late phase}: The quasi-spherical symmetry is broken, and a high accretion rate and strong feedback are maintained simultaneously. The ionized region is surrounded by cold dense gas and its shape evolves randomly depending on the dominant direction of the outflows. 

\item The mean accretion rate in the late phase increases as a function metallicity for $Z \le 0.1Z_\odot$. However, it drops to below the level for the case pristine gas when the metallicity approaches the solar abundance. 
\end{itemize}

Our numerical simulations suggest that dust contamination by star formation can play a critical role in the growth of seed BHs in the early universe. When the ISM is metal-poor ($Z \le 0.1 Z_\sun$), the growth rate is enhanced by a dust ambient medium, suggesting that the growth of seeds is closely connected with co-eval star formation. In this sense, this finding is closely related to the previous work on the role of stellar bulges in boosting accretion rates \citep{Park:2016}. However, our results also suggest that when the ISM is sufficiently enriched near the solar metallicity, BH growth might be suppressed by strong feedback, implying that there is an optimal metallicity range for the rapid growth of seed black holes.

\begin{acknowledgements}
KHP and JHW are supported by the National Science Foundation
grant OAC-1835213 and NASA grants NNX17AG23G and 80NSSC20K0520. GC was supported by Overseas Research Fellowships of the Japan Society for the Promotion of Science (JSPS). Numerical simulations presented were performed using the open-source 
{\it Enzo} and the visualization package {\sc yt} \citep{Turk:2011}. 
This work used the HIVE cluster, which is supported by the National Science 
Foundation under grant number 1828187. All the simulations presented here were 
supported in part through research cyberinfrastructure resources and services 
provided by the Partnership for an Advanced Computing Environment (PACE) at the 
Georgia Institute of Technology, Atlanta, Georgia, USA.
\end{acknowledgements}

\bibliographystyle{aasjournal}
\bibliography{park_bh}

\label{lastpage}

\end{document}